\documentclass[12pt,preprint]{aastex}

\usepackage{epsfig}
\usepackage{rotating}

\begin{document}

\title{Off-axis emission from relativistic plasma flows}

\author{E.V. Derishev\altaffilmark{1}, F.A. Aharonian\altaffilmark{2},
Vl.V. Kocharovsky\altaffilmark{1}}

\affil{$^1$~Institute of Applied Physics, 46 Ulyanov st., 603950
Nizhny Novgorod, Russia\\$^2$~Max-Planck-Institut f\"ur
Kernphysik, Saupfercheckweg 1, D-69117 Heidelberg, Germany}

\begin{abstract}

We show that there is no universal law describing how the spectra
and luminosity of synchrotron and inverse Compton radiation from
relativistic jets change with increasing observation angle.
Instead, the physics of particle acceleration leaves pronounced
imprints in the observed spectra and allows for a freedom in
numerous modifications of them. The impact of these effects is the
largest for high-energy radiation and depends on the details of
particle acceleration mechanism(s), what can be used to
discriminate between different models. Generally, the beam
patterns of relativistic jets in GeV-TeV spectral domain are much
wider than the inverse Lorentz factor. The off-axis emission in
this energy range appear to be brighter, have much harder spectra
and a much higher cut-off frequency compared to the values derived
from Doppler boosting considerations alone.

The implications include the possibility to explain high-latitude
unidentified EGRET sources as off-axis but otherwise typical
relativistic-jet sources, such as blazars, and the prediction of
GeV-TeV afterglow from transient jet sources, such as Gamma-Ray
Bursts. We also discuss the phenomenon of  beam-pattern broadening
in application to neutrino emission.

\end{abstract}


\maketitle

\section{Introduction}

Fast-moving plasma outflows are core elements in models of bright
and rapidly variable astrophysical sources, such as Active
Galactic Nuclei (AGNs), Gamma-Ray Bursts (GRBs), and microquasars
(see, for example, Urry \& Padovani 1995 for a review on AGNs,
Zhang \& M\'{e}sz\'{a}ros 2004; Piran 2005 for reviews on GRBs).
Relativistic flows offer a simple solution to the gamma-ray
transparency problem for compact objects: thanks to the Lorentz
boosting, variability timescales -- as seen in the lab frame --
decrease, the energies of individual photons and bolometric
brightness increase, whereas the pair-production opacity can be
made smaller than unity for a source of any size.

Relativistic flows can be formed by hot plasma left behind a
relativistic shock or exist in the form of jets. In either case
the question about properties of emission from these flows
naturally divides into two. Firstly, one has to know the
distribution of radiating particles in the plasma comoving frame
-- a problem far from complete solution as it depends on details
of the acceleration mechanism. Secondly, the resulting photon
field needs to be recalculated for the lab frame, i.e.,
Lorentz-transformed from the comoving frame. This is regarded as a
routine and obvious procedure and usually receives little
attention.

So far, all calculations were made under silent assumption that
the radiation is isotropic in the comoving frame. In many
situations the aforementioned isotropy is a good guess, as is
discussed in more detail in the following section. In this paper,
however, we would like to emphasize that this assumption is
model-dependent, and that highly relativistic jets and shock waves
violate it in many situations. Abandoning the isotropy assumption
has a profound effect on the predicted spectra and timing of the
astrophysical sources with relativistic plasma flows (hereafter we
will call them jets for short, that does not mean we exclude
shocks from consideration).

\begin{figure}[h]
\label{frames} \psfig{file=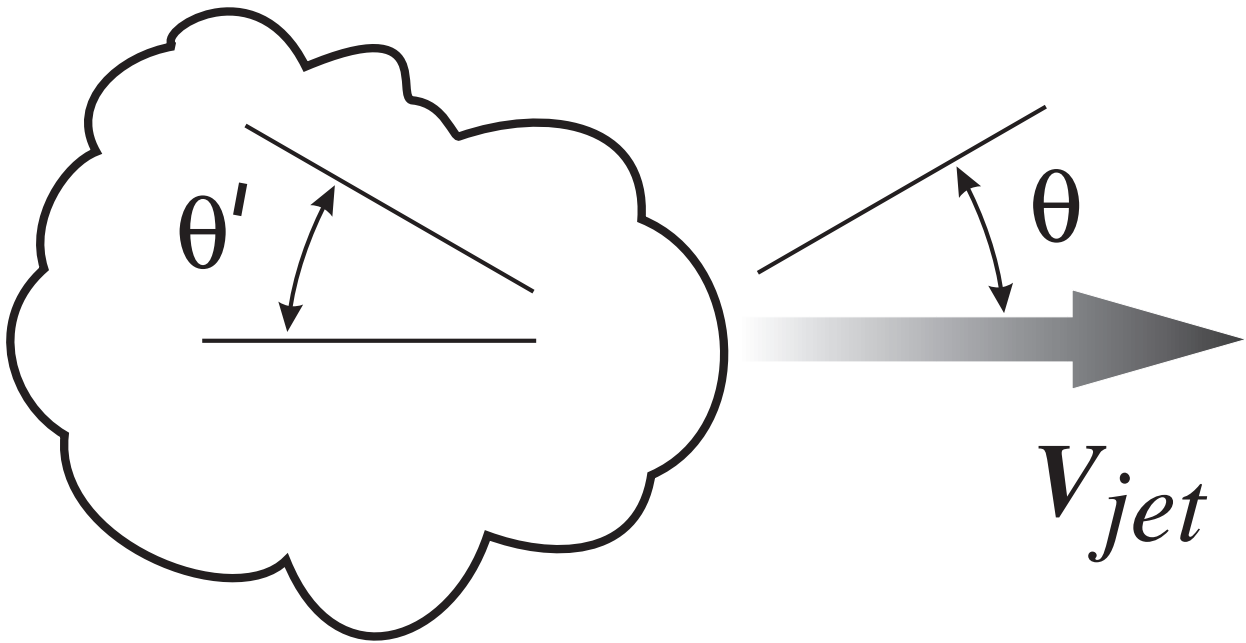,width=1.0\columnwidth}
\caption{Definition of angles in the jet comoving frame
($\theta^{\prime}$) and in the lab frame ($\theta$).}
\end{figure}

For convenience, let us remind the Lorentz transformations for the
quantities to be used in this paper: emission and observation
angles ($\theta^{\prime}$ and $\theta$, respectively),
frequencies, and intensities, i.e., energy flux per unit solid
angle and per unit frequency. (Prime stands for the jet-frame in
the case we need to introduce both the lab-frame and the jet-frame
quantities.) The reference frames are defined as shown in Fig.
\ref{frames}; note that the polar axes in the comoving and
laboratory frames are directed opposite to each other. Then, we
have the following relations
\begin{eqnarray}
\label{angle}   \mbox{for angles:} & & \cos \theta = \frac{\beta -
\cos \theta^{\prime}} {1 - \beta \cos \theta^{\prime}} \qquad
\mbox{or} \qquad \tan{\frac{\theta}{2}}
\tan{\frac{\theta^{\prime}}{2}} = \frac{1}{\Gamma (1+\beta)}\, , \\
\label{freqc}   \mbox{for frequencies:} & & \nu = \Gamma \left( 1
- \beta \cos \theta^{\prime} \right) \nu^{\prime} \equiv  \delta
\nu^{\prime}\, , \\
\label{intens} \mbox{for intensities:} & & I (\nu, \theta) =
\delta^n I^{\prime} (\nu^{\prime}, \theta^{\prime})\, .
\end{eqnarray}
Here $\beta$ is the jet velocity in units of the speed of light
$c$, $\Gamma$ the jet Lorentz factor, and
\begin{equation}
\delta = \Gamma (1 - \beta \cos \theta^{\prime})= \frac{1} {\Gamma
(1 - \beta \cos \theta)} =\frac{\sin \theta^{\prime}}{\sin \theta}
\end{equation}
the Doppler factor. In the last equation, $n=2$ for a continuous
jet and $n=3$ for a relativistically moving blob (Lind \&
Blandford 1985). The flux density per unit frequency, $F_{\nu}$,
changes with the observation angle as intensity.

Throughout the paper we will often use the asymptotic form of
Eqs.~(\ref{angle}), (\ref{freqc}) and (\ref{intens}) for the case,
where both the observation angle and the emission angle are small
($\theta, \theta^{\prime} \ll 1$). In terms of observation angle,
this condition is equivalent to $\Gamma^{-1} \ll \theta \ll 1$. In
this case
\begin{equation} \label{transf}
\delta \simeq \frac{\Gamma \theta^{\prime\, 2}}{2} \simeq
\frac{2}{\Gamma \theta^{2}}, \qquad \theta \theta^{\prime} \simeq
\frac{2}{\Gamma}.
\end{equation}
In this paper we consider only uniform jets, i.e., those with
Lorentz factor independent on direction and distance from the
origin. However, we use this simplification only when considering
particular implications in order to present conclusions in a clear
form.

The paper is organized as follows. First, we discuss the
conditions causing strong off-axis emission from relativistic jets
(Sect.~2), then we analyze motion and radiation of particles
responsible for this emission (Sect.~3 and 4). After discussing
some general results in Sect.~5, we move on to consider
implications for particular astrophysical objects (Sect.~6).

\section{Conditions for anisotropic emission in the jet frame}

Almost in every model of compact objects with relativistic flows
the observed radiation comes from particles, which are themselves
relativistic in the comoving frame (such a flow is sometimes
called a hot jet). Photons produced by a relativistic particle
continue to stream in the direction of the particle's motion,
hence the photon field in the jet frame has the same degree of
anisotropy as the particle distribution does. There always exists
a continuous anisotropic supply of relativistic particles --
those, entering the flow from surrounding medium through the shock
front or the shear layer at the jet's boundary. If their
scattering length $\ell_s$ is shorter than or comparable to their
radiation length $\ell_r$ (both are measured in the jet frame),
then the resulting emission can be treated as isotropic.
Otherwise, the radiation is essentially anisotropic in the jet
frame; in the extreme case it is directed opposite to the velocity
of the flow and confined within a narrow cone with the opening
angle $\sim \Gamma^{-1}$. Let's define the critical energy $E_{\rm
cr}^{\prime}$ so that $\ell_s (E_{\rm cr}^{\prime}) = \ell_r
(E_{\rm cr}^{\prime})$. Since $\ell_s/\ell_r$ is a monotonically
growing function of energy\footnote{The scattering length of a
particle in the magnetic field grows linearly proportional to the
particle's energy or faster, whereas the radiation length either
decreases or grows no faster than $E/\ln(E)$.}, the super-critical
particles (those with energies $E>E_{\rm cr}$) do not get
isotropized and produce anisotropic emission.

One can notice that the very nature of diffusive acceleration
implies that $\ell_s \lesssim \ell_r$. Indeed, the diffusive
acceleration proceeds through multiple passages of an accelerated
particle from the jet to surrounding medium and back, and in each
round the particle's energy increases by a factor $\sim 2$, even
in the case of relativistic shock (e.g., Achterberg et~al. 2001).
The particle should be able to preserve at least a half of its
energy over the scattering length to keep accelerating, so that
the condition $\ell_s \lesssim \ell_r$ is automatically satisfied,
providing seemingly firm ground to the isotropy assumption.

However, energetic particles emerge in the jet in several other
ways as well. They can be accelerated by preceding shocks and
survive till the shocked gas slows down to sub-relativistic
speeds, or be the secondary electrons from inelastic interactions
of higher-energy protons, or be injected in the surrounding medium
by the jet itself, as in the case of $e^{-}e^{+}$-pair loading
ahead of GRB shocks (e.g., Madau \& Thompson 2000). Also,
super-critical particles can be produced via non-diffusive
converter acceleration mechanism (Derishev et~al. 2003; Stern
2003), in which the energy gain per shock crossing is not limited
to the factor $\sim 2$. The super-critical particles do not get
isotropized by definition, and hence they contribute to the
anisotropy of photon field in the jet frame. Although both
electrons and protons can be super-critical, the case of electrons
is of much larger importance since the acceleration of protons is
almost always limited by their diffusive escape of by the
accelerator's lifetime rather than by radiative losses.

It is safe to claim that the emission from relativistic jets is
partly contributed by super-critical particles, though how large
is this part is the question to be considered separately for
various sources. Anyway, smallness of the anisotropic part in the
particle distribution does not mean its contribution to the
observed emission is negligible in all spectral domains and at all
viewing angles.

\section{Motion of super-critical particles and their emission}

Suppose that super-critical particles are injected in the jet with
highly anisotropic angular distribution, having the opening angle
$\phi_0$ (in the comoving frame) and elongated counter to the
jet's velocity. Over their radiation length, the particles get
deflected by an angle $\phi$, which is a decreasing function of
energy. The characteristic width of the beam pattern in the jet
frame ($\theta_b$) equals to that of the particle distribution:
\begin{equation}
\label{width} \theta_b^{\prime} (\varepsilon) = \left\{
\begin{array}{lll}
1& \mbox{for} &  \varepsilon < 1\\
\phi (\varepsilon)& \mbox{for} & 1 < \varepsilon < \varepsilon_{\rm cr\, 2}\\
\phi_0& \mbox{for} & \varepsilon > \varepsilon_{\rm cr\, 2}\,\, .
\end{array}
\right.
\end{equation}
To simplify the notations here and below, we introduce the
following dimensionless variables: $\varepsilon$ is the particle's
energy in units of the critical energy $E_{\rm cr}^{\prime}$ (so
that $\varepsilon_{\rm cr} \equiv 1$), $x$ the distance in units
$\ell_{s} (E_{\rm cr}^{\prime})$, which is the scattering length
at the critical energy. The second critical energy
$\varepsilon_{\rm cr\, 2}$ is defined as the energy of particles
whose r.m.s. deflection angle equals to the initial width of the
particle distribution: $\phi (\varepsilon_{\rm cr\, 2}) = \phi_0$.

Equation (\ref{width}) has simple physical meaning. The
sub-critical particles ($\varepsilon < 1$) have enough time to get
fully isotropized, whereas for super-critical ones the width of
the distribution function is equal to their r.m.s. deflection
angle $\phi$, unless $\phi < \phi_0$. Above the second critical
energy $\varepsilon_{\rm cr\, 2}$, the particles loose energy
before being significantly deflected, and the distribution
function preserves its intrinsic width $\phi_0$. The latter can be
as small as $\phi_0 =\Gamma^{-1}$ (if the distribution is
isotropic in the lab frame): for example, the converter
acceleration mechanism essentially implies isotropisation of
accelerated particles each time when they get into the surrounding
medium (Derishev et~al. 2003).

Typically, the plasma in relativistic flows is collisionless,
i.e., the only scattering mechanism is deflection of charged
particles by the magnetic field. For a broad class of the
magnetic-field geometries the dependence of the r.m.s. deflection
angle on the distance travelled by the particle allows the
following representation (for small deflection angles and
approximately constant energy):
\begin{equation} \label{defl}
    \phi =\frac{x^{p}}{\varepsilon}.
\end{equation}
This comprises two widely used limiting cases: small-angle
scattering in the purely chaotic magnetic field ($p=1/2$) and
regular deflection in the quasi-uniform magnetic field ($p=1$),
which corresponds to Bohm-like diffusion. Values in the range $1/2
< p < 1$ cover intermediate situations, including scattering in
the turbulent magnetic field with power-law power spectrum.

The number of possible energy loss mechanisms, on the other hand,
is large. Each of them implies different dependence of the energy
loss rate on the particle energy and (in general) on the
propagation angle. Moreover, in different energy domains prevalent
and negligible mechanisms can swap their roles. To keep the paper
reasonably concise, we focus our analysis on three representative
cases, marked below by the dominant energy loss channel.

{\bf Synchrotron or inverse-Compton (including self-Compton)
radiation in the Thomson regime.} In this case the energy loss
rate is proportional to the square of the particle's energy:
\begin{equation} \label{loss1}
\frac{{\rm d \varepsilon}}{{\rm d} x} = -\varepsilon^2 \qquad
\Rightarrow \qquad x_r = \frac{1}{\varepsilon}\, ,
\end{equation}
where $x_r$ is the normalized radiation length. Substituting Eq.
(\ref{loss1}) into Eq. (\ref{defl}) we find that the beam-pattern
width is
\begin{equation}\label{width1}
   \theta_b^{\prime} = \varepsilon^{-(p+1)}
\end{equation}
for particles in the energy range $1 < \varepsilon <
\varepsilon_{\rm cr\, 2}$.

The typical frequency of emitted photons is $\nu^{\prime} =
\varepsilon^2 \nu_{\rm cr}^{\prime}$, and the second critical
energy is $\displaystyle \varepsilon_{\rm cr\, 2} =
\phi_0^{-\frac{1}{p+1}}$.

{\bf Inverse-Compton radiation in the Klein-Nishina regime.} If
the spectrum of radiation being comptonized is a power-law,
$F_{\nu}^{\prime} \propto \nu^{\prime\, q}$, where $-1 < q<1$,
then
\begin{equation} \label{loss2}
\frac{{\rm d \varepsilon}}{{\rm d} x} = -\varepsilon^{1-q} \qquad
\Rightarrow \qquad x_r = \varepsilon^{q}\, .
\end{equation}
For positive $q$ the radiation length increases with increasing
energy, but anyway slower than the scattering length. Thus,
substitution of Eq. (\ref{loss2}) into Eq. (\ref{defl}) gives the
beam-pattern width
\begin{equation}\label{width2}
   \theta_b^{\prime} = \varepsilon^{pq-1},
\end{equation}
which exhibits regular (monotonically decreasing) dependence on
the particle energy.

The typical frequency of emitted photons is $\nu^{\prime} =
\varepsilon \nu_{\rm cr}^{\prime}$, and the second critical energy
is $\displaystyle \varepsilon_{\rm cr\, 2} =
\phi_0^{-\frac{1}{1-pq}}$.

{\bf Comptonization of external (isotropic in the lab frame)
radiation in the Thomson regime.} In this case the radiating
particles loose their energy through interaction with photons,
whose distribution is highly anisotropic in the jet frame. Because
of this, the energy loss rate depends on the particle's
propagation angle, but -- for an ultrarelativistic particle -- it
does not depend on the particular choice of reference frame.
Making use of this fact, we calculate the energy loss rate in the
lab frame, where it is simply $\dot{E} \propto - E^2$, and find
from Eq. (\ref{transf}) that $E \propto \phi^2 \varepsilon$. We
get
\begin{equation} \label{loss3}
\frac{{\rm d \varepsilon}}{{\rm d} x} = - \phi^4 \varepsilon^2
\qquad \Rightarrow \qquad x_r = \frac{1}{\phi^4 \varepsilon}\, ,
\end{equation}
where the proportionality coefficient is uniquely defined by the
condition $x_r (1) =1$. Solving Eqs. (\ref{defl}) and
(\ref{loss3}) for the beam-pattern width gives
\begin{equation}\label{width3}
   \theta_b^{\prime} = \varepsilon^{-\frac{p+1}{4p+1}}\, .
\end{equation}

If the external radiation has a logarithmically narrow spectrum,
then the typical frequency of emitted photons is $\displaystyle
\nu^{\prime} = \theta_b^{\prime\, 2} \varepsilon^2 \nu_{\rm
cr}^{\prime} = \varepsilon^{\frac{6p}{4p+1}} \nu_{\rm
cr}^{\prime}$. The second critical energy is $\displaystyle
\varepsilon_{\rm cr\, 2} = \phi_0^{-\frac{4p+1}{p+1}}$.

The three radiation mechanisms described above give rise to nine
qualitatively different situations, depending on what is the main
energy loss channel (this determines the beam-pattern width) and
what type of emission is observed (this determines the spectral
shape). Clearly, the physics of the off-axis emission is still
richer due to a possibility of interplay between various radiation
mechanisms.

\section{Off-axis spectra and luminosity}

In what follows, we take a power-law injection of super-critical
particles, $|{\rm d} \dot{N}/{\rm d} E| \propto E^{-s}$ for
$E>E_{\rm cr}$, where $\dot{N} (E) \propto E^{1-s}$ is the
injection rate integrated over energies larger than $E$, and $s>1$
to keep the number of injected particles finite. In the jet frame,
this transforms into the power-law injection with the same index
at energies $\varepsilon
>1$ and presumably narrow angular distribution ($\Gamma^{-1} \leq
\phi_0 \ll 1$). Generalization for the case of arbitrary injection
is straightforward, though cumbersome; we skip it for the sake of
brevity.

Because of their large scattering length, the super-critical
particles have practically no chance to leave the jet. Instead,
they loose energy and form a cooling distribution ${\rm d} N/ {\rm
d} \varepsilon = \dot{N} (\varepsilon)/ \dot{\varepsilon}_l$,
where $\dot{\varepsilon}_l$ is the total energy loss rate, mainly
associated with the dominant radiation mechanism. We also
introduce $\dot{\varepsilon}_e$ -- the energy loss rate via the
mechanism that produces the observed emission. The two can be the
same, in which case $\dot{\varepsilon}_l \simeq
\dot{\varepsilon}_e$, otherwise $|\dot{\varepsilon}_l| >
|\dot{\varepsilon}_e|$.

For a cooling distribution, the angle-averaged spectrum in the
comoving frame can be derived in the standard way:
\begin{equation} \label{av-bright}
\left< F^{\prime}_{\nu} (\nu^{\prime}) \right> \propto
\dot{\varepsilon}_e\, \frac{{\rm d} N}{{\rm d} \varepsilon} \left(
\frac{{\rm d} \nu^{\prime}}{{\rm d} \varepsilon} \right)^{-1}
\propto \frac{\dot{\varepsilon}_e}{\dot{\varepsilon}_l}\,\,
\nu^{\prime\, \frac{2-s-x}{x}}\, .
\end{equation}
Here we assumed that emission from each particle is monochromatic
with frequency $\nu^{\prime} \propto \varepsilon^{x}$ (a caution
is necessary with this assumption, as discussed below) and the
ratio $\dot{\varepsilon}_e (\varepsilon) /\dot{\varepsilon}_l
(\varepsilon)$ is regarded as a function of the frequency,
corresponding to emission from particles of energy $\varepsilon$.

The radiation of super-critical particles is concentrated within a
narrow cone with opening angle $\theta_b^{\prime} (\varepsilon)$:
outside of this cone the radiation is virtually absent, whereas
the apparent flux density within the cone is larger than the
angle-averaged one by the factor $4\, \theta_b^{\prime\, -2}$.
Thus,
\begin{equation} \label{jet-bright}
F^{\prime}_{\nu} (\nu^{\prime}) = \frac{4}{\theta_b^{\prime\, 2}}
\left< F^{\prime}_{\nu} (\nu^{\prime}) \right> \propto \left(
\frac{\dot{\varepsilon}_e}{\theta_b^{\prime\, 2}
\dot{\varepsilon}_l} \right) \nu^{\prime\, \frac{2-s-x}{x}}
\end{equation}
for $\nu_{\rm cr}^{\prime} < \nu^{\prime} < \nu_{\rm
max}^{\prime}$, where $\nu_{\rm max}^{\prime}$ is a function of
emission angle $\theta_b^{\prime}$.

Emission observed at an arbitrary angle to the jet axis has two
components: the radiation from sub-critical particles, whose
spectrum cuts off at $\nu_{\rm cr}$, and is continued to higher
frequencies by the radiation from super-critical particles. The
true cut-off in the off-axis spectrum at $\nu_{\rm max}$ is due to
the fact that an observer, looking at a given angle $\theta$ to
the jet axis, cannot see radiation from energetic particles, whose
r.m.s. deflection angle is smaller than $2/(\Gamma \theta)$.

All the changes in the off-axis spectrum as compared to the
ordinary head-on emission are entirely due to the factor
$\theta_b^{\prime\, -2}$, which is a rising function of particles'
energy (and hence -- of frequency). Therefore, the off-axis
spectrum is always harder, and in most cases -- much harder, than
the head-on spectrum. There is a subtle point in the assumption
that each particle emits monochromatic radiation. It works well
unless the spectrum given by Eq. (\ref{jet-bright}) is harder than
the low-frequency asymptotic in the spectrum of an individual
particle. All the hard spectra actually are determined by the
low-frequency emission of the most energetic particles and have
the corresponding spectral index. Such a spectrum covers the
frequency range $\nu_{\rm cr}^{\prime} < \nu^{\prime} < \nu_{\rm
max}^{\prime}$, extending also below $\nu_{\rm cr}^{\prime}$ up to
the point, where it intersects with the (softer) spectrum of
sub-critical radiation.

To find the observed luminosity one has to take the appropriate
energy loss rates from Eqs. (\ref{loss1}), (\ref{loss2}), and
(\ref{loss3}), substitute the beam-pattern width
$\theta_b^{\prime}$ in Eq. (\ref{jet-bright}) with the
corresponding function of energy and then -- energy with
frequency, and finally apply the Lorentz transformations given by
Eqs. (\ref{freqc}) and (\ref{intens}). So, an observer in the lab
frame, whose line of sight makes a small angle $1/\Gamma \ll
\theta \ll 1$ with the jet axis, sees the following spectrum:
\begin{equation} \label{lab-bright}
F_{\nu} (\nu, \theta) \propto \nu^{\alpha} \qquad \mbox{for}
\qquad \nu_{\rm cr} < \nu < \nu_{\rm max}\, ,
\end{equation}
where both the critical frequency, $\nu_{\rm cr} = \delta \nu_{\rm
cr}^{\prime}$, and the cut-off frequency, $\nu_{\rm max} = \delta
\nu_{\rm max}^{\prime}$, are functions of the observation angle.
The values of spectral index $\alpha$, as well as the ratio
$\nu_{\rm max}/ \nu_{\rm cr}$, can be found in Tab. \ref{summary},
where the summary on the resulting spectra for nine different
cases is presented.

\begin{sidewaystable}[h]
  \centering
\renewcommand{\arraystretch}{0}

\begin{tabular}{|c|@{}c|c|c|c|}
  \cline{3-5}
\multicolumn{2}{c|}{\strut } & \multicolumn{3}{c|}{Dominant energy loss mechanism} \\
  \cline{1-1} \cline{3-5}
\strut \parbox{2.2cm}{Observed\\ emission} && Synchrotron or IC & IC in the KN Regime & External Compton \\
  \cline{1-1} \cline{3-5}
  \multicolumn{5}{c}{\rule{0cm}{0.1cm}}\\
  \cline{1-1} \cline{3-5}

\parbox{2.2cm}{Synchrotron\\ or IC} &&

\parbox{6.3cm}{
\begin{eqnarray*}
  \alpha &=& p+1 -\frac{s}{2} \\
  k &=& \frac{2}{p+1} \\
  a &=& -\frac{2p}{p+1} \\
  b &=& \frac{2}{p+1}-\frac{s}{p+1}-2n
\end{eqnarray*}}&

\parbox{7.5cm}{
\begin{eqnarray*}
\alpha &=& \frac{3}{2} - \left( p- \frac{1}{2} \right)q -
\frac{s}{2} \\
  k &=& \frac{2}{1-pq} \\
  a &=& \frac{2pq}{1-pq} \\
  b &=& \frac{3+q-s}{1-pq}-2n
\end{eqnarray*}}&

\parbox{7cm}{
\begin{eqnarray*}
\alpha &=& \frac{3(p+1)}{4p+1} -\frac{s}{2} \\
  k &=& 2+\frac{6p}{p+1} \\
  a &=& \frac{6p}{p+1} \\
  b &=& \frac{6(2p+1)}{p+1}-\frac{4p+1}{p+1}s-2n
\end{eqnarray*}}

\\   \cline{1-1} \cline{3-5}
\parbox{2.2cm}{IC in the\\ KN Regime} &&

\parbox{6.3cm}{
\begin{eqnarray*}
\alpha &=& 2(p+1) -q -s \\
  k &=& \frac{1}{p+1} \\
  a &=& -\frac{2p+1}{p+1} \\
  b &=& \frac{1-q-s}{p+1}-2n
\end{eqnarray*}}&

\parbox{7cm}{
\begin{eqnarray*}
\alpha &=& 3 -2pq -s \\
  k &=& \frac{1}{1-pq} \\
  a &=& \frac{2pq-1}{1-pq} \\
  b &=& \frac{2-s}{1-pq}-2n
\end{eqnarray*}}&

\parbox{6.5cm}{
\begin{eqnarray*}
\alpha &=& \frac{6(p+1)}{4p+1} -q -s \\
  k &=& 1+\frac{3p}{p+1} \\
  a &=& \frac{3p}{p+1}-1 \\
  b &=& \frac{3p}{p+1}-\frac{4p+1}{p+1}(q+s)+5-2n
\end{eqnarray*}}

\\ \cline{1-1} \cline{3-5}
\parbox{2.2cm}{External\\ Compton} &&

\parbox{6.3cm}{
\begin{eqnarray*}
\alpha &=& \frac{4}{3} (p+1) - \frac{4p+1}{6p} s \\
  k &=& \frac{6p}{(p+1)(4p+1)} \\
  a &=& -\frac{2p}{p+1}-\frac{2}{4p+1} \\
  b &=& \frac{4p-2}{(p+1)(4p+1)}-\frac{s}{p+1}-2n
\end{eqnarray*}}&

\parbox{7cm}{
\begin{eqnarray*}
\alpha &=& \frac{4p+1}{6p} (3+q-2pq-s) - \frac{p+1}{3p} \\
  k &=& \frac{6p}{(1-pq)(4p+1)} \\
  a &=& \frac{6p}{(1-pq)(4p+1)}-2  \\
  b &=& \frac{4p-2}{(1-pq)(4p+1)}+\frac{1+q-s}{1-pq}-2n
\end{eqnarray*}}&

\parbox{6.5cm}{
\begin{eqnarray*}
\alpha &=& \frac{2(p+1)}{3p} - \frac{4p+1}{6p} s \\
  k &=& \frac{6p}{p+1} \\
  a &=& \frac{4p-2}{p+1} \\
  b &=& \frac{4p+1}{p+1}(2-s)-2n
\end{eqnarray*}}

\\  \cline{1-1} \cline{3-5}
\end{tabular}

\caption{The summary on indices, which describe the spectrum of
off-axis emission, $F_{\nu} \propto \nu^{\alpha}$, and its extent
in frequency, $\nu_{\rm max}/\nu_{\rm cr} = \left( \Gamma \theta
/2 \right)^k$. The table also presents the angular dependence of
the cut-off frequency and luminosity at the peak, $\nu_{\rm max}
\propto \theta^a$ and $L_{\rm peak} \propto \theta^b$,
respectively. The viewing angle is in the range $\Gamma^{-1} \ll
\theta \ll 1$. For the details on evaluation of these indices see
text. }

  \label{summary}
\end{sidewaystable}

Another important aspect of the off-axis emission is the way its
appearance changes with the observation angle. It can be
characterized by dependence of the cut-off frequency on the
viewing angle,
\begin{equation}\label{index-a}
\nu_{\rm max} (\theta) = \delta (\theta) \nu_{\rm max}^{\prime}
(\theta^{\prime}) = \frac{\nu_{\rm max}}{\nu_{\rm cr}}
\frac{\delta (\theta)}{\delta (0)} \nu_{\rm cr} (0) \propto
\theta^a\, ,
\end{equation}
and by the jet luminosity taken at the cut-off, $L_{\rm peak}
(\theta) = \nu_{\rm max} F_{\nu} (\nu_{\rm max}, \theta)$,
\begin{equation}\label{index-b}
L_{\rm peak} (\theta) = \left( \frac{\nu_{\rm max}}{\nu_{\rm cr}}
\right)^{\alpha+1} \delta^{n+1} (\theta) \nu_{\rm cr}^{\prime}
(\theta^{\prime}) F^{\prime}_{\nu} (\nu_{\rm cr}^{\prime},
\theta^{\prime}) = \left( \frac{\nu_{\rm max}}{\nu_{\rm cr}}
\right)^{\alpha+1} \left( \frac{\delta (\theta)}{\delta (0)}
\right)^{n+1} L_{\rm peak} (0) \propto \theta^b\, .
\end{equation}
The values of indices $a$ and $b$ are presented in Tab.
\ref{summary}.

So far, we considered only narrow jets, i.e., those having opening
angle smaller than or of the order of $\Gamma^{-1}$. For a wide
relativistic flow, one needs to integrate over observation angles,
which are different for different portions of the flow.

\section{Discussion}

The off-axis emission is intrinsically high-energy phenomenon. In
the case of Bohm diffusion, for instance, the critical energy for
electrons, whose acceleration is limited by the synchrotron
losses, is
\begin{equation}
E_{\rm cr}^{\prime} = \frac{3}{2} \frac{ \left( m_e c^2
\right)^2}{\sqrt{e^3 B^{\prime}}}\, ,
\end{equation}
and the associated cut-off frequency of their synchrotron emission
is at
\begin{equation}
\nu_{\rm cr}^{\prime}  \simeq \frac{0.5}{\pi} \frac{e
B^{\prime}}{m_e c} \left( \frac{E_{\rm cr}^{\prime}}{m_e c^2}
\right)^2, \qquad h \nu_{\rm cr}^{\prime} \simeq \frac{9}{4}
\frac{m_e c^2}{\alpha_f} \simeq 310\, m_e c^2\, ,
\end{equation}
where $\alpha_f$ is the fine structure constant. In the observer's
frame the cut-off is blueshifted to GeV range. However, a
diffusion faster that the Bohm one results is a smaller cut-off
frequency. For example, in the case of random small-angle
scattering
\begin{equation}
\label{maxsy} h \nu_{\rm cr}^{\prime} \simeq  \left(
\frac{\ell_{\rm c}}{r_{\rm g0}} \right)^{2/3} \left(
\frac{\alpha_f B^{\prime}}{B_{\rm cr}} \right)^{1/3} \frac{m_e
c^2}{\alpha_f}\, ,
\end{equation}
where $r_{\rm g0} = m_e c^2/eB^{\prime}$ is the ``cold''
gyroradius, $\ell_{\rm c}$ the correlation length of the magnetic
field, and $B_{\rm cr} \simeq 4.4 \times 10^{13}$~G. The factor
$\ell_{\rm c}/r_{\rm g0}$ can be as small as unity (if $\ell_{\rm
c} < r_{\rm g0}$, then electrons radiate in the undulator regime
and their cut-off frequency increases with decreasing
magnetic-field scale), and inverse Compton losses further decrease
the value of $\nu_{\rm cr}^{\prime}$. In the case of GRBs, where
$B^{\prime} \sim 10^5 - 10^6$~G, we find from Eq.~(\ref{maxsy})
that the cut-off can be located at just few MeV, so that the
radiation above the peak in GRB spectra can be interpreted as
off-axis synchrotron emission.

An important factor to be kept in mind when considering the
off-axis emission is two-photon pair production. Absorption of
high-energy photons in this process rapidly makes a source opaque
with increase of the viewing angle, effectively limiting its
maximum value. There is one particular situation, where
interference from the two-photon absorption is always important:
it is inverse Compton off-axis emission in the Klein-Nishina
regime in the case, where it is the dominant radiation mechanism.
Indeed, the off-axis emission implies fast cooling of radiating
electrons, i.e., the probability that they interact with target
photons is close to unity. The same is true for the comptonized
high-energy photons, since the cross-sections for electron-photon
and photon-photon interactions are of the same order of magnitude
in the Klein-Nishina limit.

As easy to see from Tab.~\ref{summary}, a spectral index of the
off-axis emission, as a rule, exceeds $-1$. In fact, this is
always the case as long as the injection spectrum is hard ($s<2$).
As the injection gets softer, there appear exceptions. The first
to break this rule (what happens at any $s>2$) is IC emission in
the Klein-Nishina regime for the case, where it dominates energy
losses, the spectrum of comptonized radiation is $F_{\nu} \propto
\nu$, and the magnetic field is quasi-uniform ($p=1$). The above
preconditions, taken together, make this situation rather
unlikely. The more common synchrotron emission, on the other hand,
is quite resistant in the hardening trend: only very soft
injection with $s>4$ can make its spectral index smaller than
$-1$. So, in the vast majority of situations, the luminosity at
cut-off, $L_{\rm peak}$, is roughly the same as the bolometric
luminosity of the jet.

Since it has many applications, it is interesting to discuss the
synchrotron emission in more detail. In the case where it is the
dominant radiation mechanism, the spectral index between $\nu_{\rm
cr}$ and $\nu_{\rm max}$ increases by 2 (for the Bohm-like
diffusion) or by 1.5 (for random small-angle scattering) relative
to what would be the spectral index of ordinary head-on emission.
For an injected particle distribution with indices $s<10/3$ or
$s<7/3$ (the Bohm-like diffusion and the small-angle scattering,
respectively), the resulting spectrum formally appears to be
harder than the low-frequency asymptotic for the synchrotron
emission of an individual particle. In practice, this means that
the spectrum is determined by the low-frequency emission of the
most energetic particles. The observed cut-off frequency depends
on the viewing angle as $\nu_{\rm max} \propto \theta^{-1}$ for
the Bohm-like diffusion and $\nu_{\rm max} \propto \theta^{-2/3}$
for the small-angle scattering, that is, much weaker than dictated
by the Lorentz transformations alone ($\nu_{\rm max} \propto
\theta^{-2}$).

Prevalence of the external Compton losses reverses the above
dependence and even cancels out the effect of jet dimming with
increasing viewing angle. Indeed, one finds from Tab.
\ref{summary} that the cut-off frequency increases with viewing
angle as $\nu_{\rm max} \propto \theta^3$ or $\nu_{\rm max}
\propto \theta^2$ for the Bohm-like diffusion and the small-angle
scattering, respectively. Under a widely used assumption that the
particle injection function has spectral index $s=2$, the peak
luminosity $L_{\rm peak}$ of a continuous jet appears to be
independent on the viewing angle for any $p$. Moreover, a hard
injection with the index $s<2$ makes an off-axis jet to appear
brighter than when it is viewed head-on.

An intermediate situation takes place in the case where
self-Compton radiation in the Klein-Nishina regime dominates the
energy losses. Here the cut-off frequency may increase or decrease
with the viewing angle, depending on whether the spectral index
$q$ of the radiation being comptonized is positive or negative.

The off-axis radiation is not necessarily electromagnetic in its
nature; for instance, it can be neutrino emission. The only
practical source of neutrinos in relativistic jets is the decay of
charged pions, which are produced in photo-pionic reactions or in
inelastic collisions of nucleons. To be precise, we note that
decaying charged pions give muons plus only one half of the total
number of muon neutrinos and anti-neutrinos. Another half and all
of the electron neutrinos and anti-neutrinos come from subsequent
decays of secondary muons. In this way, neutrinos are born
alongside with energetic photons, electrons, and positrons, which
altogether carry about a half of the energy of decaying pions.
This argument apparently leads to the conclusion that the neutrino
luminosity of a relativistic jet is at most as large as its
electromagnetic luminosity.

Once again, the common wisdom does not work with the off-axis
emission. A situation is possible, where the jet is opaque for the
high-energy photons, which therefore get reprocessed through
electromagnetic cascade, producing isotropic in the jet-comoving
frame soft electromagnetic radiation. The latter is strongly
beamed in the laboratory frame due to jet's motion. Neutrinos, on
the other hand, preserve their initial anisotropy in the comoving
frame and can be efficiently emitted at larger angles to the jet
axis. When observed at large viewing angles, such a jet looks as
an over-efficient neutrino source.

Since the off-axis neutrino emission can originate only from
anisotropic angular distribution of the parent pions (muons), it
requires substantially anisotropic -- in jet frame -- source of
energetic nucleons. It is possible if acceleration of protons is
radiative-loss limited or if there is a neutron component in the
jet (Derishev, Kocharovsky, \& Kocharovsky 1999), which moves with
the Lorentz factor different from that of the bulk matter. In
either case the decay length of pions (muons) must be less than
their scattering length not to let them isotropize. In terms of
diffusion coefficient ${\cal D} (E)$, which depends only on the
particle's energy in the ultra-relativistic limit, this condition
means
\begin{equation}
\label{D-iso} {\cal D} (E) > {\cal D}_{\rm i} (E) = \frac{1}{3}
\frac{t_{\rm i}}{m_{\rm i}}\, E\, ,
\end{equation}
where the index $\rm i$ stands either for charged pions ($\pi$) or
muons ($\mu$), $t_{\pi} \simeq 2.6 \times 10^{-8}$~s and  $t_{\mu}
\simeq 2.2 \times 10^{-6}$~s are their lifetimes, $m_{\pi}$ and
$m_{\mu}$ their masses. If ${\cal D}_{\pi} < {\cal D} < {\cal
D}_{\mu}$, then only a half of muon neutrinos contribute to the
off-axis emission, whereas the beam-pattern for electron neutrinos
and the rest of muon ones is similar to that of ordinary emission.

In the case of Bohm diffusion, Eq.~(\ref{D-iso}) translates simply
into an upper limit for the magnetic field strength:
\begin{equation}
\label{B-iso} B < B_{\rm i} =\frac{m_{\rm i} c}{e\, t_{\rm i}}\, .
\end{equation}
Here $B_{\pi} \simeq 600$~G and $B_{\mu} \simeq 5$~G, so that the
above condition is true for any potential neutrino sources except
arguably for the GRB internal shocks, where the diffusion should
be orders of magnitude faster than the Bohm diffusion to fulfill
Eq.~(\ref{B-iso}).

\section{Implications}

Many astrophysical sources with relativistic jets change their
appearance in presence of the off-axis emission. The difference is
negligible at low frequencies, but becomes dramatic for
high-energy photons (typically X- and gamma-rays). Unfortunately,
it is practically impossible to make definitive and unequivocal
predictions from the first principles since the properties of the
off-axis emission strongly depend on details of both radiation and
acceleration mechanisms, with uncertainties in geometry further
increasing the range of possible solutions. The problem, however,
has a silver lining from the observational perspective: the very
same diversity of unique observational signatures provides a means
to determine physical parameters in a source.

In accordance with the above note, this section is not to present
a comprehensive analysis of the properties of off-axis emission
for various sources, but rather to give an idea of what one
expects in typical situations, that is done below using primarily
GRBs as a representative example.

GRBs are a complex phenomenon (see, e.g., [] and [] for a review),
which can be decomposed into qualitatively different
prompt-emission and afterglow phases. During the prompt phase,
which lasts from a fraction of a second to few hundred seconds,
GRBs usually have highly irregular lightcurves and relatively hard
emission. The afterglow is characterized by gradually decaying
smooth lightcurve with occasional rises and regular softening of
the emission.

In the following discussion we assume for definiteness that peaks
in observed GRB spectra correspond to transition from sub-critical
to super-critical radiation regimes, so that the radiation above
the peak is mainly due to off-axis emission. Such an
interpretation implies that both sub- and super-critical particles
form a single distribution. This is possible if the radiating
electrons are secondary particles from inelastic interactions of
high-energy protons, or produced via a non-diffusive (for example,
converter) acceleration mechanism.

The prompt emission of GRBs is thought to be the synchrotron
radiation originating from a succession of internal shocks, i.e.,
those developing within the fireball at a distance of the order of
$D \sim 10^{12}$~cm from the central engine. Radiation from a
large number of such shocks contributes to observed flux at any
moment of time; in effect, they can be treated as a continuous
jet. One can imagine two situations: a jet, whose opening angle
$\theta_0$ is smaller than the viewing angle $\theta$, and the
opposite case of small viewing angle, $\theta < \theta_0$.

The timing properties of the prompt emission are similar in both
cases. Since the Doppler factor for off-axis jets is smaller, the
variability timescale must be longer. However, the total duration
of a burst $t_{\rm GRB}$ is not affected: it is determined by the
lifetime of central engine, at least as far as geometrical delay
for light propagation is smaller than the lifetime, i.e.,
$\theta^2 D/2 c \la t_{\rm GRB}$. Even for short bursts, the
latter condition corresponds to relatively large viewing angles
$\theta \la 0.1$.

For the case of small viewing angle, observed spectra are affected
in two ways. The off-axis emission from edge portions of the jet
(those propagating at angles much larger than $1/\Gamma$ to the
line of sight) can contribute to: (1) the bolometric luminosity
and (2) a high-energy tail above the cut-off frequency. For the
synchrotron-self-Compton emission, no matter whether the
synchrotron or inverse Compton losses are dominant, the peak
luminosity $L_{\rm peak} (\theta)$ normally \footnote{The opposite
requires either hard injection with $s<2$ or the low-frequency
asymptotic in the spectrum of comptonized radiation harder than
$F_{\nu} \propto \nu^{1/3}$, that is a source of emission other
than the synchrotron.} decreases with increase of the viewing
angle faster than $\theta^{-2}$, making the first effect
negligible. On the contrary, if the jet looses energy mostly to
external Compton radiation, then the edge portions of the jet
dominate the overall bolometric luminosity as far as $s<14/5$,
that is, for any reasonable injection. Although prevalence of
external Compton losses in the jet's radiative balance or an
injection with the index $s<2$ are not favored by current GRB
theories, we conclude that the edge portions of the jet cannot
safely be ignored even when calculating the bolometric luminosity.

If the cut-off frequency $\nu_{\rm max}$ increases with increasing
viewing angle, then the off-axis emission from edges of a wide jet
significantly changes the observed (composite) spectrum, causing a
high-energy tail to appear instead of an exponential cut-off.
Parts of the jet viewed at different angles contribute to this
tail with luminosities $\propto \theta L_{\rm peak} (\theta)
\propto \theta^{b+1}$, concentrated mostly around frequency $\nu
(\theta) \simeq \nu_{\rm max} (\theta) \propto \theta^{a}$. The
envelope of individual contributions gives the power-law tail:
\begin{equation}
\nu F_{\nu} \propto \nu \left( \theta L_{\rm peak}\, \frac{{\rm
d}\, \theta}{{\rm d}\, \nu} \right) \propto \nu^{\frac{2+b}{a}}\,
,
\end{equation}
where $\theta \propto \nu^{1/a}$ and $a>0$. As follows from Tab.
\ref{summary}, the condition $a>0$ can be satisfied in a
consistent synchrotron-self-Compton model, for example if
comptonization proceeds in the Klein-Nishina regime and the
spectral index of comptonized radiation is positive (i.e., $q>0$),
so that emergence of the power-law tail should be considered a
common phenomenon.

In the case of large viewing angle, $\theta \gg \theta_0$, every
portion of a jet moves at approximately the same angle to the line
of sight, so that the situation is in almost every respect
equivalent to the case of narrow jet, which was considered in
Sect.~4. The only correction to be made is to take into account
that the jet subtends an angle, which is much larger than
$1/\Gamma$. For an idealized (uniform with sharp edges) jet the
corrected dependence of luminosity on the viewing angle is
\begin{equation}
L_{\rm peak} (\theta) = (\Gamma \theta)^{b} (\Gamma \theta_0)^2
L_{\rm peak} (0)\, ,
\end{equation}
where $b<-2$ and $\theta, \theta_0 \gg \Gamma^{-1}$. As the
viewing angle exceeds the opening angle of the jet, the bolometric
luminosity drops by a factor $\sim (\Gamma \theta_0)^{-2-b}$.

Taking into consideration the off-axis emission, it is interesting
to discuss a possibility that the X-ray flashes (XRFs) are normal
GRBs, whose jets are not pointing to the observer. The peak energy
in XRF spectra is smaller than in GRB spectra, implying that $a<0$
and, consequently (see Tab.~\ref{summary}), that $b<-3$.
Therefore, in the case of idealized jet, the XRFs and GRBs are
members of separate source populations, whose average brightness
differs by a factor $\Gamma \theta_0 \ll 1$ or smaller. Apart from
this difference in brightness, the XRF spectra in their low-energy
(below the peak) part should possess the intrinsic feature of
off-axis emission -- a paucity of soft photons.

Finally, let us discuss of a tempting possibility to explain
unusually weak GRBs as off-axis jet. Unlike XRFs, weak GRBs have
luminosities many orders of magnitude smaller than their normal
counterparts, but radiate in the same spectral range. The off-axis
synchrotron emission can account for the properties of weak GRBs
if their main radiative mechanism is self-Compton in the
Klein-Nishina regime, and their low-frequency spectral index $q$
is close to zero. Indeed, in this case the cut-off frequency is
nearly independent on the viewing angle, whereas the observed
luminosity can drop as fast as $\theta^{-3}$.

Unlike the prompt GRB emission, the afterglow comes from a single
blast wave, which forms when the GRB ejecta plunge into
surrounding interstellar gas and which in most cases can be
approximated by a thin spherical shell. Despite the simple
geometry, dynamics of this blast wave is complicated by various
factors, such as inhomogeneous external medium, formation of
multiple sub-shocks, late energy injection, etc., which are beyond
the scope of this paper. However, a principal part of the problem
-- namely, obtaining the Green function for a radiating spherical
shell -- can be formulated in model-independent terms. Physically,
the Green function $G(R,t,\nu)$ is the spectral flux density,
measured by a distant observer as a function of time, provided the
radiation comes from an instantaneous release of unit energy in a
spherical shell of radius $R$ expanding with velocity $v(R)$. By
definition, $\displaystyle \int G(R,t,\nu)\, {\rm d}\nu\, {\rm d}t
=1$ and $G \equiv 0$ for any $t<0$. The spectrum of any thin blast
wave can be represented as
\begin{equation}
\label{lightcurve} F_{\nu} (t,\nu) = \int_0^{\infty} \lambda (R)\,
G(R,t-t_e,\nu)\, {\rm d}R \, ,
\end{equation}
where $\lambda (R)$ is the energy lost for radiation per unit
distance, and $\displaystyle t_e (R) = \int_0^{R} \frac{{\rm d}R}{
v(R)} -\frac{R}{c}$\quad the time, as measured by the distant
observer, it takes for the blast wave to expand to the radius $R$.
If necessary, Eq.~(\ref{lightcurve}) includes another integration
to take into account the radial structure of the blast wave.

To find the Green function, we note that the area of spherical
segment that comes into the observer's view during the time
interval ${\rm d}t$ is equal to $2\pi R c {\rm d}t$. This segment
moves at an angle $\theta(t)=\arccos (1-ct/R)$ to the line of
sight and its contribution to the detected fluence is proportional
to $\theta^{k(\alpha+1)}\delta^{n+1} {\rm d}t \propto
\delta^{-b/2} {\rm d}t$, where the index $n$ (see
Eq.~(\ref{intens}) for definition) is equal to 2. Indeed,
physically an element of the blast wave is a blob, whose
luminosity scales with $n=3$, and whose apparent lifetime is
proportional to $\delta^{-1}$, so that its fluence scales with
$n=2$. So, we obtain
\begin{equation}
G(R,t,\nu) = \left[ \frac{2}{b+2} \frac{R}{\beta c} \left\{
(1+\beta)^{1+b/2} - (1-\beta)^{1+b/2} \right\} \right]^{-1}
\Theta(t)\, \Theta \left(\frac{2R}{c} -t \right) \left( 1-\beta +
\beta \frac{ct}{R} \right)^{b/2} f(R,\nu,\theta) \, ,
\end{equation}
where $\Theta(t)$ is the step function, $f(R,\nu,\theta)$ the
spectral energy distribution, normalized to unity, and the factor
in square brackets ensures that the Green function as a whole is
normalized to unity.

Since we are interested in the ultra-relativistic case, where
$\beta \rightarrow 1$, it is convenient to use the approximate
expression for the Green function,
\begin{equation}
\label{gf} G(R,t,\nu) \simeq \left[ - (b+2) \Gamma^2 \frac{c}{R}
\right] \Theta(t) \left( 1 + 2 \Gamma^2 \frac{ct}{R} \right)^{b/2}
f(R,\nu,\theta) \, ,
\end{equation}
which is also valid for a blast wave with finite angular extent as
long as the opening angle is much larger than $\Gamma^{-1}$.

An important thing to learn from Eq.~(\ref{gf}) is that a
radiating shell fades away rather gradually after the shock has
passed it, producing what may be called a geometrical, or
retarded, afterglow. Due to the geometrical delay, the retarded
emission from early afterglow coexists in time with ordinary
emission from late afterglow, and its bolometric luminosity
asymptotically decreases as $t^{b/2}$. In absence of the off-axis
emission, $b=-6$ and the luminosity of geometrical afterglow
rapidly decays to a level, indiscernible against the much brighter
ordinary afterglow. For the off-axis emission, the index $b$ is
typically in the range $-4<b<-2$, that corresponds to decay rate
of the geometrical afterglow between $t^{-2}$ and $t^{-1}$. For
comparison: the bolometric luminosity of ordinary afterglow
behaves as $t^{-3/2}$ and $t^{-12/7}$ for adiabatic and fully
radiative shocks, respectively, propagating into uniform medium,
or as $t^{-1}$ and $t^{-4/3}$ if the shocks propagate into a wind
with density profile $\rho \propto R^{-2}$.

The properties of off-axis emission, which are discussed above in
application to GRBs, show up also in AGNs (and microquasars, as
far as they can be considered a scaled-down version of AGNs).
Thus, we limit our analysis of AGNs to only one specific point --
the observational bias against detection of jets pointing away
from the line of sight.

With present-day telescopes AGN surveys are sensitivity-limited,
i.e., we detect only those, whose apparent brightness is above
certain threshold. Let us suppose that the bolometric luminosity
changes with the viewing angle as $\theta^{-b}$ and the sources
are uniformly distributed in space. Then the number of detectable
sources decreases with viewing angle as
\begin{equation}
N (\theta) \propto \theta^{\frac{4-3b}{2}} \, .
\end{equation}
In absence of the off-axis emission the bolometric luminosity
decreases with the viewing angle as $\theta^{-6}$, so that $N
(\theta) \propto \theta^{-7}$. Among hundreds of known blazars one
can hardly expect to find even a single source with $\theta >
2/\Gamma$, in accordance with the existing observational data. On
the other hand, the radiogalaxies are observed with randomly
directed jets, that comes at no surprise since the radio-emission
is not beamed. The off-axis emission is much less beamed than the
ordinary radiation from AGN inner jets and -- in this respect --
resembles radio-emission, though it occupies the opposite end of
electromagnetic spectrum. The number of detectable off-axis
sources increases with increasing index $b$ and they dominate the
entire source population for any $b>-4/3$.

It turns out that a situation, typical for blazars (the radiative
losses are mostly due to inverse Compton scattering of external
photons), provides also an extreme example of the off-axis
synchrotron emission, with apparent luminosity almost independent
on the viewing angle. It means that the majority of blazars, which
are not detected at present because of large inclination of their
jets to the line of sight, will show up when observed in the right
spectral range. For the synchrotron off-axis emission from MeV
blazars the most favorable (from the observational point of view)
spectral domain is around 100 MeV, within the operational range of
GLAST and AGILE. The inverse Compton component of the off-axis
emission can be detected with modern ground-based Cherenkov
telescopes, which are sensitive to photons down to 30-100 GeV.
Some of these off-axis blazarz may have already been detected by
EGRET as unidentified high-latitude sources.

\section{Conclusions}

A number of processes lead to generation of super-critical
particles in relativistic flows. Having the scattering length of
the order of the radiation length or exceeding it, these particles
do not isotropize upon entering the relativistic flow and radiate
their energy while preserving a certain degree of anisotropy. This
anisotropy counteracts the beaming, which results from the Lorentz
boost, so that the emission produced in such a way has a wider
beam pattern in the laboratory frame than that of sub-critical
particles and can be called the off-axis emission. The properties
of the off-axis emission under various conditions are summarized
in Table~\ref{summary}. Among many implications considered in this
paper, the following are of major importance from the
observational point of view.

The jet sources, which are observed off-axis owing to the effect
of beam pattern broadening should exhibit very hard spectra.
Indeed, for the super-critical particles the r.m.s. deflection
angle (and hence the width of the beam pattern) is a function of
their energy. An observer situated at a large angle to the jet
axis effectively sees the particle distribution devoid of its
low-energy part, whose emission can only be seen at smaller
viewing angles. Therefore, the off-axis emission is the hardest
possible -- in most cases it is essentially as hard as the
spectrum of an individual particle. An off-axis jet (for example,
AGN, GRB, or microquasar) is likely to be a source of gamma-ray
radiation above several MeV and up to TeV range without any bright
X-ray or optical counterpart. Apparently, the off-axis AGNs can
account at least for some of unidentified extragalactic EGRET
sources.

The space-borne gamma-ray telescopes with wide field of view, such
as GLAST and AGILE, have the greatest chance to detect synchrotron
radiation from off-axis jets. The inverse Compton component of the
off-axis emission can be detected by ground-based Cherenkov
telescopes. However, they have very limited surveying
capabilities, so that the best observing strategy in search of the
off-axis emission would be to look at known AGNs whose jet are
pointing away from the line of sight.

In transient sources (for example, GRBs) broader beam pattern
means larger geometrical delay, which is proportional to the
square of angle between the jet axis and the line of sight. The
luminosity of in the retarded off-axis emission decays rather
slowly in time, allowing observations in GeV-TeV range when the
prompt emission is over. Moreover, if the temporal index of
geometrical afterglow is larger than -3/2 (that is right in the
middle of the typical range), then the integral signal-to-noise
ratio continually grows with observing time as long as the
off-axis emission is present. This opens an interesting
possibility for observation of GRBs with ground-based Cherenkov
telescopes, which normally have to slow response to catch the
prompt radiation. A serendipitous discovery of orphan GRB
afterglows in the TeV range is also possible, because the beam
pattern is broader for high-energy photons. It should be noted
that there is observational evidence for delayed GRB emission, at
least in the case of GRB940217 (Hurley et al., 1994), which can be
interpreted as geometrical afterglow due to the off-axis emission.

The off-axis emission is intrinsically high-energy phenomenon and
in some cases it may experience two-photon absorption within the
source, especially at large viewing angles. In opaque sources the
electromagnetic radiation from super-critical particles is
reprocessed through the electromagnetic cascade, looses its
identity and becomes collimated, but the neutrino signal from them
still comes out. Such jets, when viewed off-axis, appear
over-efficient neutrino sources, where the ratio of neutrino
luminosity to the electromagnetic one can be almost arbitrarily
large. The off-axis neutrinos can be detected by the next
generation of cubic-kilometer scale high-energy neutrino
detectors, and may provide unique information on the details and
relative importance of various particle acceleration processes.

\section{Acknowledgments}

E.V. Derishev acknowledges the support from the President of the
Russian Federation Program for Support of Young Scientists (grant
no. MK-2752.2005.2). This work was also supported by the RFBR
grants no. 05-02-17525 and 04-02-16987, the President of the
Russian Federation Program for Support of Leading Scientific
Schools (grant no. NSh-4588.2006.2), and the program "Origin and
Evolution of Stars and Galaxies" of the Presidium of the Russian
Academy of Science.


\begin{thebibliography}{}
\bibitem{kirk} Achterberg, A., Gallant, Y.A., Kirk, J.G.,
Guthmann, A.W. 2001, MNRAS, 328, 393
\bibitem{neutr} Derishev, E.V., Kocharovsky, V.V., \& Kocharovsky,
Vl. V. 1999, ApJ, 521, 640
\bibitem{ours} Derishev, E.V., Aharonian, F.A., Kocharovsky, V.V.,
Kocharovsky, Vl.V. 2003, Phys. Rev. D, 68, 043003
\bibitem{GeVdelay} Hurley, K. et al. 1994, Nature, 372, 652
\bibitem{beam} Lind, K.R., \& Blandford, R.D. 1985, ApJ, 295, 358
\bibitem{pairs} Madau, P., \& Thompson, C. 2000, ApJ, 534,
239
\bibitem{GRBrew2} Piran, T. 2005, Rev. Mod. Phys., 76, 1143
\bibitem{stern} Stern, B.E. 2003, MNRAS, 352, L35
\bibitem{AGNrew} Urry, C.M., \& Padovani, P. 1995, PASP, 107, 803
\bibitem{GRBrew1} Zhang, B., \& M\'{e}sz\'{a}ros, P. 2004, IJMPA, 19, 2385
\end{thebibliography}
\end{document}